\documentclass[pre,aps,nofootinbib,twocolumn]{revtex4}%
\usepackage{amssymb}
\usepackage{graphicx}
\usepackage{amsmath}
\usepackage{amsfonts}%
\providecommand{\U}[1]{\protect\rule{.1in}{.1in}}
\begin{document}
\author{ $^{1,2}$M. L. Kuli\'{c}, $^{3}$O. V. Dolgov}
\title{The electron-phonon interaction with forward scattering peak is a relevant
approach to high $T_{c}$ superconductivity in $FeSe$ films on $SrTiO_{3}$ and
$TiO_{2}$ }

\address{$^{1}$Institute for Theoretical Physics, Goethe-University D-60438
Frankfurt am Main, Germany \\
$^{2}$Institute of Physics, Pregrevica 118, 11080 Belgrade (Zemun), Serbia\\
$^{3}$Max-Planck-Institut f\"{u}r Festk\"{o}rperphysik,70569
Stuttgart, Germany}

\begin{abstract}
The theory of the electron-phonon interaction ($EPI$) with strong forward
scattering peak ($FSP$) in an extreme delta-peak limit \cite{Kulic-Zeyher1-2}%
-\cite{Dan-Dol-Kul-Oud} is recently applied in
\cite{Lee-Interfacial-supplement}-\cite{Johnston1-2} for the explanation of
high $T_{c}(\sim100$ $K)$ in a monolayer $FeSe$ grown on $SrTiO_{3}$
\cite{Lee-Interfacial-supplement} and $TiO_{2}$ \cite{Rebec-FeSe-TiO2}
substrates. The $EPI$ is due to a long-range dipolar electric field created by
the high-energy oxygen vibrations ($\Omega\sim90$ $meV$) at the interface
\cite{Lee-Interfacial-supplement}-\cite{Johnston1-2}. We show that in leading
order (with respect to $T_{c0}/\Omega$) the mean-field critical temperature
$T_{c0}=\left\langle V_{epi}(q)\right\rangle _{q}/4)$ $\sim$ $(aq_{c}%
)^{2}V_{epi}(0)$ and the gap $\Delta_{0}=2T_{c0\text{ }}$ are due to an
interplay between the maximal $EPI$ pairing potential $V_{epi}(0)$ and the
$FSP$-width $q_{c}$. For $T_{c0}\sim100$ $K$ one has $\Delta_{0}\sim16$ $meV$
in a satisfactory agreement with $ARPES$ experiments. We find that in leading
order $T_{c0}$ is \textit{\ mass-independent }and a very small oxygen isotope
effect is expected in next to leading order. In clean systems $T_{c0}$ for
$s$-wave and $d$-wave pairing is degenerate but both are affected by
non-magnetic impurities, which are \textit{pair-weakening} in the $s$-channel
and \textit{pair-breaking} in the $d$-channel.

The self-energy and replica bands at $T=0$ and at the Fermi surface are
calculated and compared with the corresponding results at $T>0$
\cite{Johnston1-2}. The $EPI$ coupling constant $\lambda_{m}=\left\langle
V_{epi}(q)\right\rangle _{q}/2\Omega$, which enters the self-energy
$\Sigma(k,\omega)$, is \textit{mass-dependent} ($M^{1/2}$) which at
$\omega(\ll\Omega)$ makes the slope of $\ \Sigma(k,\omega)(\approx-\lambda
_{m}\omega)$ and the replica intensities $A_{i}(\sim\lambda_{m})$
\textit{mass-dependent}. This result, overlooked in the literature, is
contrary to the prediction of its mass-independence in the standard
Migdal-Eliashberg theory for $EPI$. The small oxygen isotope effect in
$T_{c0}$ and pronounced isotope effect in $\Sigma(k,\omega)$ and $ARPES$
spectra $A_{i}$ of replica bands in $FeSe$ films on $SrTiO_{3}$ and $TiO_{2}$
is a smoking-gun experiment for testing an applicability of the $EPI-FSP$
theory to these systems. The $EPI-FSP$ theory predicts a large number of
low-laying pairing states (above the ground state) thus causing internal pair
fluctuations. The latter reduce $T_{c0}$ additionally, by creating a pseudogap
state for $T_{c}<T<T_{c0}$.

Possibilities to increase $T_{c0}$, by designing novel structures are
discussed in the framework of the $EPI-FSP$ theory.

\end{abstract}
\date{\today}
\maketitle

\section{Introduction}

The scientific race in reaching high temperature superconductivity ($HTSC$)
started by the famous Ginzburg's proposal of an \textit{excitonic mechanism of
pairing} \textit{in metallic-semiconducting sandwich-structures}
\cite{Ginzburg-excitons}. In such a system an electron from the metal tunnels
into the semiconducting material and virtually excites high-energy exciton,
which is absorbed by another electron, thus making an effective attractive
interaction and Cooper pairing. However, this beautiful idea has not been
realized experimentally until now. In that sense V. L. Ginzburg founded a
theoretical group of outstanding and talented physicists, who studied at that
time almost all imaginable pairing mechanisms. In this group an important role
has played the Ginzburg's collaborator E. G. Maksimov, who was an "inveterate
enemy" of almost all other mechanisms of pairing in $HTSC$ but for the
electron-phonon one - see his arguments in \cite{Dolg-Kirzh-Maks}. It seems
that the recent discovery of superconductivity in a $Fe$-based material made
of one monolayer film of the iron-selenide $FeSe$ grown on the $SrTiO_{3}$
substrate - further called $1ML$ $FeSe/SrTiO_{3}$, with the critical
temperature $T_{c}\sim(50-100)$ $K$ \cite{Wang QJ-1-Lee JJ}, as well as grown
on the rutile $TiO_{2}$ (100) substrate with $T_{c}\sim65$ $K$
\cite{Rebec-FeSe-TiO2} - further called $1ML$ $FeSe/TiO_{2}$, in some sense
reconciles the credence of these two outstanding physicists. Namely, $HTSC$ is
realized in a sandwich-structure but the pairing is due to an high-energy
($\sim90-100$ $meV$) \textit{oxygen optical phonon}. This (experimental)
discovery will certainly revive discussions on the role of the electron-phonon
interaction ($EPI$) in $HTSC$\ cuprates and in bulk materials of the
$Fe$-pnictides (with the basic unit $Fe-As$) and Fe-chalcogenides (with the
basic unit $Fe-Se$ or $Te$, $S$). As a digression, we point out that after the
discovery of high $T_{c}$ in $Fe$-pnictides a non-phononic pairing mechanism
was proposed immediately, which is due to: $(i$) nesting properties of the
electron- and hole-Fermi surfaces and ($ii$) an enhanced (due to ($i$)) spin
exchange interaction ($SFI$) between electrons and holes
\cite{Hirschfeld-review}. This mechanism is called the \textit{nesting SFI
pairing}. However, the discovery of alkaline iron selenides $K_{x}%
Fe_{2-y}Se_{2}$ with $T_{c}\sim30$ $K$, and intercalated compounds
$Li_{x}(C_{2}H_{8}N_{2})Fe_{2-y}Se_{2}$, $Li_{x}(NH_{2})_{y}(NH_{3}%
)_{1-y}Fe_{2}Se_{2}$, which contain \textit{only electron-like Fermi
surfaces}, rules out the nesting pairing mechanism as a common pairing
mechanism in Fe-based materials. In order to overcome this inadequacy of the
$SFI$ nesting mechanism a pure phenomenological "strong coupling" $SFI$
pairing is proposed in the framework of the so called $J_{1}-J_{2}$
Heisenberg-like Hamiltonian, which may describe the $s$-wave
superconductivity, too. However, this approach is questionable since the LDA
calculations cannot be mapped onto a Heisenberg model and there is a need to
introduce further terms in form of biquadratic exchange
\cite{Hirschfeld-review}. It is interesting, that immediately after the
discovery of high $T_{c}$ in pnictides the electron-phonon pairing mechanism
was rather uncritically discarded. This attitude was exclusively based on the
$LDA$ band structure calculations of the electron-phonon coupling constant
\cite{Boeri-Dolgov}, which in this approach turns out to be rather small
$\lambda<0.2$, thus giving $T_{c}<1$ $K$.

In the past there were only few publications trying to argue that the $EPI$
pairing mechanism is an important (pairing) ingredient in the $Fe$-based
superconductors \cite{Kulic-Dolgov-Drechsler}-\cite{Kulic-Ginzburg Conf-2012}.
One of the theoretical arguments for it, may be ilustrated in the case of
2-band superconductivity. In the weak-coupling limit $T_{c}$ is given by
$T_{c}=1.2\omega_{c}\exp\{-1/\lambda_{\max}\},$ where $\lambda_{\max}%
=(\lambda_{11}+\lambda_{22}+\sqrt{(\lambda_{11}-\lambda_{22})^{2}%
+4\lambda_{12}\lambda_{21}})/2$. In the nesting $SFI$ pairing mechanism one
assumes a dominance of the repulsive inter-band pairing ( $\lambda_{12}%
$,$\lambda_{21}<0$), i.e. $\left\vert \lambda_{12},\lambda_{21}\right\vert
\gg\left\vert \lambda_{11},\lambda_{22}\right\vert $. Since the intra-band
pairing depends on $\lambda_{ii}=\lambda_{ii}^{epi}-\mu_{ii}^{\ast}$, where
$\lambda_{ii}^{epi}$ is the intra-band $EPI$ coupling constant and $\mu
_{ii}^{\ast}>0$ is an screened intra-band Coulomb repulsion, then in order to
maximize $T_{c}$ the intra-band $EPI$ coupling it is wishful that
$\lambda_{ii}^{epi}$ at least compensate negative effects of $\mu_{ii}^{\ast}$
(on $T_{c}$), i.e. $\lambda_{ii}^{epi}\geq\mu_{ii}^{\ast}$. Since in a narrow
band one expects rather large screened Coulomb repulsion $\mu_{ii}^{\ast}$
($\sim0.2$) then the intra-band $EPI$ coupling should be also appreciable.
Moreover, from the experimental side the Raman measurements in Fe-pnictides
\cite{Keimer-A1g} give strong evidence for a large phonon line-width of some
$A_{1g}$ modes (where the $As$ vibration along the c-axis dominates). They are
almost $10$ times larger than the $LDA$ band structure calculations predict.
In \cite{Kulic-Haghighirad} a model was proposed where high electronic
polarizability of $As$ ($\alpha_{As^{3-}}\sim12$ \AA $^{3}$) ions screens the
Hubbard repulsion and also give rise to a strong $EPI$ with A$_{1g}$ (mainly
$As$) modes. An appreciable $As$ isotope effect in $T_{c0}$ was proposed in
\cite{Kulic-Haghighirad}, where the stable $^{75}As$ should be replaced by
unstable $^{73}As$ - with the life-time of $80$ days, quite enough for
performing relevant experiments. The situation is similar with $Fe-Se$
compounds, where an appreciable $EPI$ is expected, since $^{78}Se$ is also
highly polarizable ($\alpha_{Se^{2-}}\sim7.5$ \AA $^{3}$) and can be replaced
by a long-living $^{73}Se$ isotope - the half-time $120$ days. Unfortunately
these experiments were never performed.

We end up this digression by paying attention to some known facts, that the
$LDA$ band structure calculations are unreliable in treating most high $T_{c}$
superconductors, since as a rule $LDA$ \textit{underestimates non-local
exchange-correlation effects and overestimates charge screening effects} -
both effects contribute significantly to the $EPI$ coupling constant. As a
result, $LDA$ strongly underestimates the $EPI$ coupling in a number of
superconductors, especially in those near a metal-isolator transition. The
classical examples for this claim are: (\textit{i}) the $(BaK)BiO_{3\text{ }}%
$superconductor with $T_{c}>30$ $K$ which is $K$-doped from the parent
isolating state $BaKBiO_{3}$. Here, $LDA$ predicts $\lambda<0.3$ and
$T_{c}\sim1$ $K$, while the theories with an appropriate non-local
exchange-correlation potential \cite{Yin-Kutepov-Kotliar-PRB} predict
$\lambda_{epi}\approx1$ and $T_{c}\sim31$ $K$; (\textit{ii}) The high
temperature superconductors, for instance $YBaCu_{3}O_{7}$ with $T_{c}\sim100$
$K$, whose parent compound $YBaCu_{3}O_{6}$ is the Mott-insulator
\cite{Kulic-PhysRep}, \cite{Ma-Ku-Do-Advances}.

After this digression we consider the main subject of the paper - \textit{the
role of the }$EPI$\textit{\ with forward scattering peak (}$FSP$\textit{) in
pairing mechanism} of the $1ML$ $FeSe/SrTiO_{3}$ (and also $1ML$
$FeSe/TiO_{2}$) superconductor(s) with high critical temperatures $T_{c}%
\sim(50-100)$ $K$. In that respect, numerous experiments on $1ML$
$FeSe/SrTiO_{3}$ (and also on$1ML$ $FeSe/TiO_{2}$), combined with the fact
that the $FeSe$ film on the graphene substrate has rather small $T_{c}%
\approx8$ $K$ (like in the bulk $FeSe$), give strong evidence that interface
effects, due to $SrTiO_{3}$ (and $TiO_{2}$), are most probably responsible for
high $T_{c}$. It turns out, that the most important results in $1ML$
$FeSe/SrTiO_{3}$ (and also $1ML$ $FeSe/TiO_{2}$), related to the existence of
quasi-particle \textit{replica bands} - which are identical to the main
quasiparticle band \cite{Lee-Interfacial-supplement}, \cite{Rebec-FeSe-TiO2},
\cite{Replica-bands}, can be coherently described by the $EPI-FSP$%
\textit{\ }theory. This approach was proposed in seminal papers
\cite{Lee-Interfacial-supplement}-\cite{Johnston1-2}. The beauty of these
papers lies in the fact that they have recognized sharp replica bands in the
$\acute{A}RPES$ spectra and related them to a sharp forward scattering peak in
the $EPI$. (This is a very good example for a constructive cooperation of
experimentalists and theoreticians.) Let us mention, that the $EPI-FSP$ theory
was first studied in a connection with $HTSC$ cuprates \cite{Kulic-Zeyher1-2},
while the extreme case of the $EPI-FSP$ pairing mechanism with delta-peak is
elaborated in \cite{Dan-Dol-Kul-Oud} - see a review in \cite{Kulic-PhysRep}.
Physically, this (in some sense exotic) interaction means that in some
specific materials (for instance in cuprates and in $1ML$ $FeSe/SrTiO_{3}$)
electron pairs exchange virtual phonons with small (transfer) momenta
$q<q_{c}\ll k_{F}$ only, and as a result the effective pairing potential
becomes long-ranged in real space \cite{Kulic-PhysRep}. It turns out that this
kind of pairing can in some cases give rise to higher $T_{c}$ than in the
standard (Migdal-Eliashberg) $BCS$-like theory. Namely, in the $EPI-FSP$
pairing mechanism one has $T_{c}^{(FSP)}=\left\langle V_{epi}(q)\right\rangle
_{q}/4\sim\lambda_{epi}^{(FSP)}/N(E_{F})$ \cite{Dan-Dol-Kul-Oud} - see below,
instead of the $BCS$ dependence $T_{c}^{(BCS)}\sim\Omega e^{-1/\lambda
_{epi}^{(BCS)}}$. Here, $\lambda_{epi}^{(FSP)}$ and $\lambda_{epi}^{(BCS)}$
are the corresponding \textit{mass-independent} $EPI$ coupling constants,
where $\Omega$ - is the phonon energy, $N(E_{F})$ - the electronic density of
states (per spin) at the Fermi surface.\ So, even for small $\lambda
_{epi}^{(FSP)}\ll\lambda_{epi}^{(BCS)}$ the case $T_{c}^{(FSP)}>T_{c}^{(BCS)}$
can be in principle realized. We inform the reader in advance, that the
$EPI-FSP$ theory predicts also that $T_{c}^{(FSP)}\sim(q_{c}/k_{F})^{d}%
V_{epi}(0)$, ($d=1,2,3$ is the dimensionality of the system), which means that
when $q_{c}\ll k_{F}$ high $T_{c}^{(FSP)}$ is hardly possible in $3D$ systems.
However, the detrimental effect of the phase-volume factor $(q_{c}/k_{F})^{d}$
on $T_{c}^{(FSP)}$ can be compensated by its linear dependence on the pairing
potential $V_{epi}(0)$. In some favorable materials this competition may lead
even to an increase of $T_{c}$. We stress that properties of the
superconductors with the $EPI-FSP$ mechanism of pairing are in many respects
very different from the standard (BCS-like) superconductors, and it is
completely justified to speak about \textit{exotic superconductors}. For
instance, the $EPI-FSP$ theory \cite{Kulic-PhysRep}-\cite{Dan-Dol-Kul-Oud}
predicts, that in superconductors with the $EPI-FSP$ pairing \textit{the
isotope effect should be small} in leading order, i.e. $\alpha\ll1/2$
\cite{Kulic-PhysRep}-\cite{Dan-Dol-Kul-Oud} - see discussion in the following.
This result is contrary to the case of the isotropic $EPI$ theory in standard
metallic superconductors, where $\alpha$ is maximal, $\alpha=1/2$ (for
$\mu^{\ast}=0$). We point out, that the $EPI-FSP$ pairing mechanism in
strongly correlated systems is rather strange in comparison with the
corresponding one in standard metals with good electronic screening, where the
large transfer momenta dominate and the pairing interaction is, therefore,
short-range. As a result, an important consequence of the $EPI-FSP$ pairing
mechanism in case of $HTSC$-cuprates is that $T_{c}$ in the $d$-wave channel
is of the same order as in the $s$-wave one. Since the residual repulsion is
larger in the $s$- than in the $d$-channel ($\mu_{d}^{\ast}\ll\mu_{s}^{\ast}$)
this result opens a door for $d$-wave pairing in $HTSC$-cuprates, in spite of
the fact of the $EPI$ dominance \cite{Kulic-Zeyher1-2}-\cite{Kulic-PhysRep}.

In the following, we study the superconductivity in $1ML$ $FeSe/SrTiO_{3}$
(and $1ML$ $FeSe/TiO_{2}$ \cite{Rebec-FeSe-TiO2}) in the framework of a
semi-microscopic model of $EPI$ first proposed in seminal papers
\cite{Lee-Interfacial-supplement}-\cite{Johnston1-2}. Namely, due to oxygen
vacancies: $(i)$ an electronic doping of the $FeSe$ monolayer is realized,
which gives rise to electronic-like bands centered at the $M$-points in the
Brillouin zone, while the top of the hole-bands are at around $60$ $meV$ below
the electronic-like Fermi surface; $(ii)$ the formed charge in the interface
orders dipoles in the nearby $TiO_{2}$ layer; $(iii)$ the free charges in
$SrTiO_{3}$ screen the dipolar field in the bulk, thus leaving the $TiO_{2}$
dipolar layer near the interface as an important source for the EPI. The
oxygen ions in the $TiO_{2}$ dipolar layer vibrate with high-energy
$\Omega\approx90$ $meV$, thus making a long-range dipolar electric field
acting on metallic electrons in the $FeSe$ monolayer. This gives rise to a
long-ranged $EPI$ \cite{Lee-Interfacial-supplement}, \cite{Johnston1-2}, which
in the momentum space gives a forward scattering peak - the $EPI-FSP$%
\textit{\ pairing mechanism}.

In this paper we make some analytical calculations in the framework of the
$EPI-FSP$ theory with a very narrow $\delta$-peak, with the width $q_{c}\ll
k_{F}$, wher $k_{F}$ is the Fermi momentum \cite{Dan-Dol-Kul-Oud}. Here, we
enumerate the obtained results, only: ($\mathbf{1}$) in leading order the
\textit{critical temperature} is linearly dependent on the pairing potential
$V_{epi}(\mathbf{q})$, i.e. $T_{c0}\approx\left\langle V_{epi}(\mathbf{q}%
)\right\rangle _{q}/4$. In order to obtain $T_{c0}\sim100$ $K$ we set the
range of semi-microscopic parameters ($\varepsilon_{\parallel}^{eff}$,
$\varepsilon_{\perp}^{eff}$, $q_{eff}$, $h_{0}$, $n_{d}$ - see below) entering
$\left\langle V_{epi}(\mathbf{q})\right\rangle _{q}$. Furthermore, since
$\left\langle V_{epi}(\mathbf{q})\right\rangle _{q}$ is independent of the the
oxygen ($O$) mass, then $T_{c0}$ is \textit{mass-independent} in leading order
with respect to $T_{c0}/\Omega$. This means, that in $1ML$ $FeSe/SrTiO_{3}$
(and $1ML$ $FeSe/TiO_{2}$) one expects very small $O$-isotope effect
($\alpha_{O}\ll1/2$). Note, in \cite{Johnston1-2} large $\alpha_{O}=1/2$ is
found; ($\mathbf{2}$) the \textit{self-energy} $\Sigma(\mathbf{k},\omega)$\ at
$T=0$ is calculated analytically which gives: ($i$) the positions and spectral
weights of the replica and quasiparticle bands at $T=0$ - all this quantities
are \textit{mass-dependent}; ($ii$) the slope of the quasiparticle self-energy
for $\omega\ll\Omega$ ($\Sigma(\omega)\approx-\lambda_{m}\omega$) is
\textit{mass-dependent}, since $\sim\lambda_{m}\sim M_{O}^{1/2}$;
($\mathbf{3}$) in the $EPI-FSP$ model (without other interactions) the
critical temperature for $s$-wave and $d$-wave pairing is degenerate, i.e.
$T_{c0}^{(s)}=T_{c0}^{(d)}$. The presence of non-magnetic impurities (with the
parameter $\Gamma=\pi n_{i}N(E_{F})u^{2}$) lifts this degeneracy. It is shown,
that even the $s$-wave pairing (in the $EPI-FSP$ pairing mechanism) is
sensitive to non-magnetic impurities, which are \textit{pair-weakening} for
it, i.e. $T_{c0}^{(s)}$ is decreased for large $\Gamma$, but never vanishes.
It is also shown that for $d$-wave pairing $T_{c0}^{(d)}$ strongly depends on
impurities, which are \textit{pair-breaking}. The curiosity is that in the
presence of non-magnetic impurities $T_{c0}^{(d)}$ in the $EPI-FSP$ pairing
mechanism is more robust than the corresponding one in the $BCS$ model;
($\mathbf{4}$) the long-range $EPI-FSP$ pairing potential in real space makes
a short-range potential in the momentum space. The latter gives rise to
numerous low-laying excitation energy (above the ground-state) of pairs, thus
leading to strong internal pair fluctuations which reduce $T_{c0}$. At
$T_{c}<T<T_{c0}$ a pseudogap behavior is expected.

The structure of the paper is following: in \textit{Section II} we calculate
the $EPI-FSP$ pairing potential as a function of semi-microscopic parameters
in the model of a dipolar layer $TiO_{2}$ with vibrations qf the oxygen ions
\cite{Lee-Interfacial-supplement}-\cite{Johnston1-2}. In \textit{Section III}
the self-energy effects, such as replica bands and their intensities at $T=0$,
are studied. The critical temperature $T_{c0}$ is calculated in
\textit{Section IV} in terms of the semi-microscopic parameters ($\varepsilon
_{\parallel}^{eff}$, $\varepsilon_{\perp}^{eff}$, $q_{eff}$, $h_{0}$, $n_{d}%
$). The range of of these parameters, for which one has $T_{c0}\sim100$ $K$,
is estimated, too. In \textit{Section V} the effect of nonmagnetic impurities
on $T_{c0}$ are studied, while the effects of internal fluctuations of Cooper
pairs are briefly discussed in \textit{Section VI}. Summary of results are
presented in \textit{Section VII}.

\section{$EPI-FSP$ pairing potential due to dipolar oxygen vibrations in the
$TiO_{2}$layer}

It is important to point out that in $1ML$ $FeSe/SrTiO_{3}$ material, with
$1$-monolayer of $FeSe$ grown on the $SrTiO_{3}$ substrate - mainly on the
$(0,0,1)$ plane, the Fermi surface in the $FeSe$ monolayer is
\textit{electron-like\ }and\textit{\ }centered at four M-points in the
Brillouin zone - see more in \cite{Rebec-FeSe-TiO2}, \cite{Sadovskii1-2}. The
absence of the (nested) hole-bands on the Fermi surface rules out all $SFI$
nesting theories of pairing. Even the pairing between an electron- and
incipient hole-band \cite{Hirschfeld-incipient} is ineffective since: ($1$) in
the $FeSe$ monolayer the top of the hole band lies below the Fermi level
around $60-80$ $meV$; ($2$) because of ($1$) the SFI coupling constant is
(much) smaller than in the nesting case. This brings into play the interface
interaction effects. The existence of \textit{sharp replica bands} in the
$ARPES$ spectra at energies of the order of optical phonons with $\Omega
\sim90$ $meV$, implies inevitably that the dominant interaction in $1ML$
$FeSe/SrTiO_{3}$ (and also in $1ML$ $FeSe/TiO_{2}$ \cite{Rebec-FeSe-TiO2}) is
due to $EPI$ with strong forward scattering peak
\cite{Lee-Interfacial-supplement}-\cite{Johnston1-2}. The physical mechanism
for $EPI-FSP$ is material dependent and the basic physical quantities such as
the width of the $FSP$, phonon frequencies and bare $EPI$ coupling can vary
significantly from material to material. For instance, in $HTSC$-cuprates the
effective $EPI-FSP$ potential $V_{epi}(q)$ is strongly renormalized by strong
correlations, which is a synonym for large repulsion of two electrons on the
$Cu$ ions - the doubly occupancy is forbidden. In that case the approximative
$q$-dependence of $V_{epi}(q)$ is given by $V_{epi}(q)\approx\lbrack
1+(q/q_{c})^{2}]^{-2}V_{0,epi}(q)$, $q_{c}\sim\delta/a$, where $V_{0,epi}(q)$
is the bare (without strong correlations) coupling constant, $\delta(\ll1)$ is
the hole concentration and $a$ is the $Cu-O$ distance \cite{Kulic-Zeyher1-2}%
-\cite{Kulic-PhysRep}. The prefactor is a vertex correction due to strong
correlations and it means a new kind of (anti)screening in strongly correlated materials.

\begin{figure}[ptb]
\includegraphics*[width=8cm]{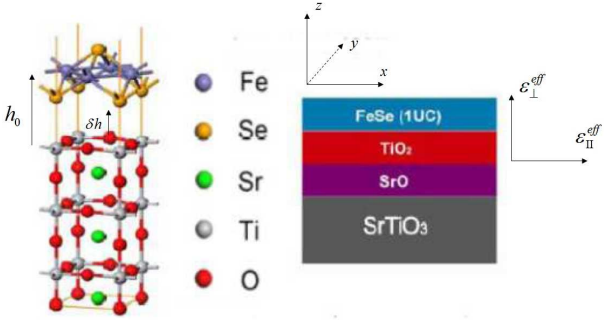}\caption{\textit{Left}: The microscopic
structure of the $FeSe/SrTiO_{3}$ interface of the $1ML$ $FeSe/SrTiO_{3}$
structure. $h_{0}$ - distance between the $FeSe$ monolayer and $TiO_{2}$
dipolar layer. $\delta h$ is amplitude of the oxygen vibration in the dipolar
layer. \textit{Right}: The perpendicular view on the $FeSe-SrTiO_{3}$
interface in the model with one $TiO_{2}$ dipolar layer at the end of the
$SrTiO_{3}$ substrate. The anisotropy of the effective dielectric constant
$varepsilon$ is shown. The similar schema holds also for the $1ML$
$FeSe/TiO_{2}$ structure}%
\label{1}%
\end{figure}

The interface in $1ML$ $FeSe/SrTiO_{3}$ can be considered as highly
anisotropic material with the parallel and perpendicular (to the $FeSe$ plane)
dielectric constants $\varepsilon_{\parallel}^{eff}\gg\varepsilon_{\perp
}^{eff}$. It is assumed \cite{Lee-Interfacial-supplement}-\cite{Johnston1-2}
that the oxygen from the $TiO_{2}$ dipolar layer - placed at height ($-h_{0}$)
from the $FeSe$ plane, vibrate and make dipolar moments $\delta p_{z}%
=q_{eff}\delta h(x,y,-h_{0})$ perpendicular to the $FeSe$ ($x-y$) plane - see
$Fig.1$. It gives rise to a dipolar electric potential $\Phi_{dip}%
(x,y,-h_{0}-\delta h)$ acting on electrons in the $FeSe$ ($x-y$ plane). Here,
$q_{eff}$ is an effective charge per dipole and $\delta h$ is the polar
(dominantly oxygen) displacement along the $z$-axis
\cite{Lee-Interfacial-supplement}. Due to some confusion in the literature on
the form of $\Phi_{dip}$ \cite{Lee-Interfacial-supplement} we recalculate it
here, in order to know its explicite dependence on the semi-microscopic
parameters $\varepsilon_{\parallel}^{eff}$, $\varepsilon_{\perp}^{eff}$,
$q_{eff}$, $h_{0}$, $n_{d}$. An elementary electrodynamics approach
\cite{Landau-Electrodynamics} gives for the dipolar potential $\Phi
_{dip}(x,y,-h_{0}-\delta h)$%

\[
\Phi_{dip}(x,y,-h_{0}-\delta h)=\frac{(\varepsilon_{\parallel}^{eff})^{1/2}%
}{(\varepsilon_{\perp}^{eff})^{3/2}}(n_{d}q_{eff}h_{0})\times
\]%
\begin{equation}%
{\displaystyle\iint}
\frac{dx^{\prime}dy^{\prime}\delta h(x^{\prime},y^{\prime},-h_{0})}{\left(
\frac{\varepsilon_{\parallel}^{eff}}{\varepsilon_{\perp}^{eff}}h_{0}%
^{2}+(x-x^{\prime})^{2}+(y-y^{\prime})^{2}\right)  ^{\frac{3}{2}}},
\label{Phi}%
\end{equation}
where $n_{d}$ is the number of the oscillating $Ti-O$ dipoles per unit $FeSe$
surface. The coefficient in front of the integral is different from that in
\cite{Lee-Interfacial-supplement} - where it is $\varepsilon_{\parallel}%
^{eff}/(\varepsilon_{\perp}^{eff})^{3/2}(q_{eff}h_{0})$ (probably typos?) and
with missed dipole density $n_{d}$. This coefficient does not fulfill the
condition $\Phi\sim\varepsilon^{-1}$ in the isotropic case, while
$Eq.$(\ref{Phi}) does. By introducing $g_{epi}(\mathbf{q})=e\Phi(\mathbf{q})$
the electron-phonon interaction Hamiltonian $H_{epi}=\sum_{\mathbf{q}}%
e\hat{\Phi}(\mathbf{q})\hat{\rho}(\mathbf{q})$ is rewritten in the form
$H_{epi}=\sum_{\mathbf{k},\mathbf{q}}g_{epi}(\mathbf{q})(\hat{b}_{\mathbf{q}%
}+\hat{b}_{-\mathbf{q}}^{\dagger})\hat{c}_{\mathbf{k}+\mathbf{q}}^{\dagger
}\hat{c}_{\mathbf{k}}$, where $\hat{b}_{-\mathbf{q}}^{\dagger},\hat
{c}_{\mathbf{k}+\mathbf{q}}^{\dagger}$ are boson and fermion creation
operators, respectively. The Fourier transformed potential $g_{epi}%
(\mathbf{q})(=(g_{0}/\sqrt{N})e^{-q/q_{c}})$ is given by
\begin{equation}
g_{epi}(\mathbf{q})=\frac{2\pi n_{d}eq_{eff}}{\varepsilon_{\perp}^{eff}}%
\sqrt{\frac{\hbar}{M\Omega N}}e^{-q/q_{c}}, \label{g-epi}%
\end{equation}
$g_{0}=(2\pi n_{d}eq_{eff}/\varepsilon_{\perp}^{eff})(\hbar/M\Omega)^{1/2}$,
$e$ is the electronic charge. Here, the screening momentum $q_{c}%
=(\varepsilon_{\perp}^{eff}/\varepsilon_{\parallel}^{eff})^{1/2}h_{0}^{-1}$
characterizes the range of the $EPI$ potential, i.e. for $q_{c}\ll k_{F}$
($k_{F}$ is the Fermi momentum) the $EPI$ is sharply peaked at $\mathbf{q}=0$
- the forward scattering peak ($FSP$), and the potential in real space is
long-ranged, while for $q_{c}\sim k_{F}$ it is short-ranged, like in the
standard $EPI$ theory. Since we are interested in the $T_{c}$ dependence on
the effective parameters $\varepsilon_{\parallel}^{eff},\varepsilon_{\perp
}^{eff},q_{eff},h_{0}$, then an explicit dependence of the potential is
important. We shall see below, that in order that this approach is applicable
to $1ML$ $FeSe/SrTiO_{3}$ (and $1ML$ $FeSe/TiO_{2}$) $\varepsilon_{\parallel
}^{eff},\varepsilon_{\perp}^{eff}$ must be very different from the bulk values
of $\varepsilon$ in the bulk $SrTiO_{3}$ - where $\varepsilon\sim500-10^{4}$,
or in the rutile $TiO_{2}$ structure where $\varepsilon<260$
\cite{Rebec-FeSe-TiO2}.

\section{Self-energy effects and $ARPES$\ replica bands}

The general self-energy $\Sigma_{epi}(\mathbf{k}_{F},\omega)$ at $T=0$ in the
extreme $FSP$ $\delta$-peak limit with the width $q_{c}\ll k_{F}$) is given by
(see Appendix)%
\begin{equation}
\Sigma_{epi}(\mathbf{k}_{F},\omega)\approx-\lambda_{m}\frac{\omega}%
{1-(\omega/\Omega)^{2}}, \label{Sigma-epi}%
\end{equation}
where $\lambda_{m}=\left\langle V_{epi}(\mathbf{q})\right\rangle _{q}/2\Omega$
is the \textit{mass-dependent coupling constant}. Here, the average $EPI$
potential is given by $\left\langle V_{epi}(\mathbf{q})\right\rangle
_{q}=Ns_{c}(2\pi)^{-2}\int d^{2}qV_{epi}(\mathbf{q},0)\approx(1/4\pi
)(aq_{c})^{2}V_{epi}^{0}$, where $s_{c}=2a^{2}$ is the surface of the $FeSe$
unit cell and $a$ is the Fe-Fe distance, and the bare pairing $EPI$ potential
is $V_{epi}^{0}=2g_{0}^{2}/\Omega$. The coupling constant $\lambda_{m}$
corresponds to $\lambda_{m}$ used in \cite{Johnston1-2}, where the self-energy
effects are studied at $T>0$. It is important to point out that $\lambda_{m}$
is (oxygen) \textit{mass-dependent}, contrary to \cite{Johnston1-2}. Since
$\left\langle V_{epi}(\mathbf{q})\right\rangle _{q}$ is
\textit{mass-independent} then $\lambda_{m}\sim\Omega^{-1}\sim M^{1/2}$. In
the following we discuss the case when $\mathbf{k}=\mathbf{k}_{F}$, i.e.
$\xi(\mathbf{k})=0$. For $\omega\ll\Omega$ one has $\Sigma_{epi}%
(\mathbf{k},\omega)=-\lambda_{m}\omega$ which means that the slope of
$\Sigma_{epi}(\mathbf{k},\omega)$ is \textit{mass-dependent}. The latter
property can be measured by $ARPES$ and thus the $EPI-FSP$ theory can be
tested. Note, that in the $EPI-FSP$ theory the critical temperature
$T_{c0}(=\left\langle V_{epi}(\mathbf{q})\right\rangle _{q}/4)$ - see details
below, is \textit{mass-independent}. Both these results are \textit{opposite
to the standard Migdal-Eliashberg theory}, where the self-energy slope is
mass-independent and $T_{c0}$ is mass-dependent.

The \textit{quasiparticle and replica bands} at $T=0$ are obtained from
$\omega-\Sigma_{epi}(\omega)=0$. In the following we make calculations at
$T=0$ and at the Fermi surface $\xi(\mathbf{k}_{F})=0$. The solutions are:
($1$) $\omega_{1}=0$ - the \textit{quasiparticle band}; ($2$) $\omega
_{2}=-\Omega\sqrt{1+\lambda_{m}}$ is the $ARPES$ \textit{replica band}; ($3$)
the inverse $ARPES$ replica band $\omega_{3}=\Omega\sqrt{1+\lambda_{m}}$. The
single-particle spectral function is $A(\mathbf{k}_{F},\omega,T=0)=\sum
_{i=1}^{3}(A_{i}/\pi)\delta(\omega-\omega_{i})$, where $A_{i}/\pi$ are the
spectral weights. For the quasiparticle band $\omega_{1}$ one obtains
$A_{1}=(1+\lambda_{m})^{-1}$, while for the replica bands at $\omega_{2}$ and
$\omega_{3}$ one has $A_{2}=A_{3}=(\lambda_{m}/2)(1+\lambda_{m})^{-1}$. The
ratio of the intensities at $T=0$ of the $\omega_{2}$ replica band and
quasiparticle band $\omega_{1}$ is given by%
\begin{equation}
\frac{A_{2}(\mathbf{k}_{F},\omega,T=0)}{A_{1}(\mathbf{k}_{F},\omega
,T=0)}=\frac{\lambda_{m}}{2}. \label{ARPES-inten}%
\end{equation}
It is necessary to mention that at finite $T(>0)$ this ratio is changed as
found in \cite{Johnston1-2}. In that case $\Sigma_{epi}^{(T)}(\mathbf{k\approx
k}_{F},\omega,T\neq0)\approx\lambda_{m}/(\omega+\Omega)$ which gives the
quasiparticle and replica band $\omega_{1}^{(T)}=\Omega(-1+\sqrt
{1+4\lambda_{m}})/2$ and $\omega_{2}^{(T)}=-\Omega(1+\sqrt{1+4\lambda_{m}})/2$
and $(A_{2}/A_{1})_{T}=\lambda_{m}$ \cite{Steve-Kulic}. This intriguing
difference of the $T=0$ and $T\neq0$ results for $(A_{2}/A_{1})$ in the
$EPI-FSP$ theory, $(A_{2}/A_{1})_{T}=2(A_{2}/A_{1})_{0}$, is due to the
sharpness of the Fermi function $n_{F}(\xi_{\mathbf{k+q}})$ entering in
$\Sigma_{epi}(\mathbf{k},\omega)\ $- see $Eq.$(\ref{Sigma-integ}) in Appendix
\cite{Steve-Kulic}.

We stress that, the $ARPES$ measurements of $A_{2}/A_{1}$ in $1ML$
$FeSe/SrTiO_{3}$ were done at finite temperatures ($T\neq0$) and in the $k=0$
point with $\xi(k=0)\sim-50$ $meV$ which gives $(A_{2}/A_{1})_{T}%
\approx0.15-0.2$ \cite{Lee-Interfacial-supplement}, \cite{Replica-bands}.
According to the theory in \cite{Lee-Interfacial-supplement},
\cite{Johnston1-2} one obtains $\lambda_{m}^{(ARPES)}\approx0.15-0.2$. Below
we show, that $\lambda_{m}$ can be also extracted from the formula
$Eq.$(\ref{Tco} ) for $T_{c0}\approx100$ $K$, which gives $\lambda
_{m}^{(T_{c0})}\approx0.18$. The latter value is in a good agreement with
$\lambda_{m}^{(ARPES)}$ from $ARPES$ \cite{Steve-Kulic}. If we put this value
in \ $Eq.$(\ref{ARPES-inten}) one obtains that at $T=0$ and at $k=k_{F}$ one
has $(A_{2}/A_{1})\sim0.1$. From this analysis we conclude that the ARPES
measurements at $k_{F}$ should give the similar ratio as at $k=0$. The
calculated $ARPES$ spectra at $T=0$ $K$ and at $k=k_{F}$ give $\Delta
\omega=\left\vert \omega_{2}-\omega_{1}\right\vert =\Omega\sqrt{1+\lambda_{m}%
}$ while the experimental value is $\Delta\omega\approx100$ $meV$, which for
$\lambda_{m}^{(ARPES)}\approx0.2$ gives the optical phonon energy of the order
of $\Omega\approx90$ $meV$.

\section{The superconducting critical temperature $T_{c0}$ and gap $\Delta
_{0}$}

In the weak coupling limit ($\lambda_{m}\ll1$) of the Eliashberg equations
with $q_{c}v_{F}<\pi T_{c0}$ ($v_{F}$ is the Fermi velocity) the linearized
gap equation (near $T_{c0}$) is given by
\begin{equation}
\Delta(\mathbf{k},\omega_{n})=T_{c0}\sum_{\mathbf{q}}\sum_{m}\frac
{V_{epi}(q,\omega_{n}-\omega_{m})\Delta(\mathbf{k}+\mathbf{q},\omega_{m}%
)}{\omega_{m}^{2}+\xi^{2}(\mathbf{k}+\mathbf{q})}, \label{Delta}%
\end{equation}
where $\omega_{n}=\pi T_{c0}(2n+1)$, $V_{epi}(\mathbf{q},\Omega_{n}%
)=g_{epi}^{2}(\mathbf{q})(2\Omega/(\Omega_{n}^{2}+\Omega^{2}))$, $\Omega
_{n}=2\pi T_{c0}\cdot n$. For $\Omega\gg\pi T_{c0}$ one has $V_{epi}%
(\mathbf{q},\Omega_{n})\approx V_{epi}(\mathbf{q},0)=2g_{epi}^{2}%
(\mathbf{q})/\hbar\Omega$. In the strong $FSP$ limit when $(q_{c}v_{F})\ll(\pi
T_{c0})^{2}$ the highest value of $\Delta(\mathbf{k},\omega_{n})$ is reached
at $\mathbf{k}=\mathbf{k}_{F}$ ($\xi(\mathbf{k}_{F})=0$ in $Eq.$(\ref{Delta}).
The solution $\Delta(\mathbf{k},\omega_{n})$ is searched in the standard
BCS-like \textit{square-well} approximation $\Delta(\mathbf{k}_{F},\omega
_{n})\approx\Delta_{0}=const$. In leading order with respect to $(T_{c0}%
/\Omega)\ll1$ one obtains $T_{c0}$ \cite{Kulic-PhysRep}-\cite{Dan-Dol-Kul-Oud}%

\begin{equation}
T_{c0}\approx\frac{1}{\pi^{2}}\left\langle V_{epi}(\mathbf{q})\right\rangle
_{q}\sum_{-\Omega/\pi T_{c0}}^{\Omega/\pi T_{c0}}\frac{1}{(2m+1)^{2}}
\label{Tco}%
\end{equation}
For $\Omega\gg\pi T_{c0}$ this gives $T_{c0}=\left\langle V_{epi}%
(\mathbf{q})\right\rangle _{q}/4\approx(1/16\pi)(aq_{c})^{2}(2g_{0}^{2}%
/\Omega)$, where $a$ is the $Fe-Fe$ distance. Note, that $T_{c0}$ is
\textit{mass-independent (}$\alpha_{O}=0$) - note $\alpha_{O}=1/2$ is found in
\cite{Johnston1-2}. The small isotope-effect can be a smoking-gun experiment
for the $EPI-FSP$ pairing mechanism in $1ML$ $FeSe/SrTiO_{3}$ (and $1ML$
$FeSe/TiO_{2}$). From $Eq.$(\ref{Delta-n-FSP}) in the Appendix it is
straightforward to obtain the \textit{energy gap} $\Delta_{0}=2T_{c0}$. Note,
that $1ML$ $FeSe/SrTiO_{3}$ (and $1ML$ $FeSe/TiO_{2}$) is a $2D$ system and
$T_{c0}\sim q_{c}{}^{2}$, while in the $d$-dimensional space one has
$T_{c0}\sim q_{c}{}^{d}$. This means that the $EPI-FSP$ mechanism of
superconductivity is more \textit{favorable in low-dimensional systems}
($d=1,2$) than in the $3D$ one. Since high $T_{c}$ cuprates are also
quasi-$2D$ systems, where strong correlations make a long-ranged $EPI$, it
means that the $EPI-FSP$ mechanism of pairing may be also operative in
cuprates \cite{Kulic-PhysRep}. Note, that in estimating some semi-microscopic
parameters we shall use as a reper-value $T_{c0}\approx100$ $K$, while in real
systems $T_{c}\sim(60-80)$ $K<T_{c0}$ is realized. However, $T_{c0}$ is the
mean-field value obtained in the Migdal-Eliashberg theory, while in $2D$
systems it is significantly reduced by the phase fluctuations - to the
Berezinski-Kosterliz-Thouless value. There is an additional reduction of
$T_{c0}$ (which might be also appreciable) in the $EPI-FSP$ systems, which is
due to internal pair-fluctuations - see discussion below.

One can estimate the coupling constant $\lambda_{m}=\left\langle
V_{epi}(\mathbf{q})\right\rangle _{q}/2\Omega$ in $1ML$ $FeSe/SrTiO_{3}$ from
the value of $T_{c0}$. Then for the reper-value $T_{c0}\sim100$ $K$ one has
$\left\langle V_{epi}(\mathbf{q})\right\rangle _{q}\approx33$ $meV$ and
$\lambda_{m}^{(T_{c0})}\approx$ $0.18$. Since, $\lambda_{m}^{(ARPES)}%
\approx\lambda_{m}^{(T_{c0})}$ the consistency of the theory is satisfactory.
Note, if one includes the wave-function renormalization effects (contained in
$Z(i\omega_{n})>1$) then in the case $(T_{c0}/\Omega)\ll1$ and for the
square-well solution $T_{c0}$ is lowered to $T_{c0}^{(Z)}=T_{c0}/Z^{2}(0)$,
where $Z(0)\approx1+\lambda_{m}$ \cite{Dan-Dol-Kul-Oud}. This means, that the
nonlinear corrections (with respect to $\lambda_{m})$ in $T_{c0}$ and
$\Delta_{0}$ \cite{Johnston1-2}, \cite{Murta} should be inevitably
renormalized by the $Z$-renormalization.

Let us estimate the parameters ($\varepsilon_{\parallel},\varepsilon_{\perp
},q_{eff},h_{0}$) which enter in $T_{c0}$. In order to reach $T_{c0}\sim100$
$K$ (and $\left\langle V_{epi}(\mathbf{q})\right\rangle _{q}=4T_{c0}\sim400$
$K\approx33$ $meV$) then for $aq_{c}\approx0.2$ and $\Omega\approx90$ $meV$
one obtains $g_{0}\approx0.7$ $eV$. Having in mind that $g_{0}=(2\pi
n_{d}eq_{eff}/\varepsilon_{\perp}^{eff})(\hbar/M\Omega)^{1/2}$ and that the
zero-motion oxygen amplitude is $(\hbar/M\Omega)^{1/2}\approx0.05$ \AA \ and
by assuming that $n_{d}\approx\alpha/s_{c}$, $s_{c}=\tilde{a}^{2}$, $\tilde
{a}=\sqrt{2}a\approx4$ \AA , $q_{eff}\sim2e$, $\alpha\gtrsim1$, then in order
to obtain $g_{0}\approx0.7$ $eV$ $\varepsilon_{\perp}^{eff}$ must be small,
i.e. $\varepsilon_{\perp}^{eff}\sim1$. Since $aq_{c}=(a/h_{0})\sqrt
{\varepsilon_{\perp}^{eff}/\varepsilon_{\parallel}^{eff}}\sim0.2$ and for
$(a/h_{0})\sim1$ it follows $\varepsilon_{\parallel}^{eff}\sim30$. Note, that
in $SrTiO_{3\text{ }}$the bulk $\varepsilon$ is large, $\varepsilon
\sim500-10^{4}$. So, if $T_{c0}$ in $1ML$ $FeSe/SrTiO_{3}$ is due solely to
the $EPI-FSP$ mechanism, then in the model where the oxygen vibrations in the
single dipolar monolayer $TiO_{2\text{ }}$are responsible for the pairing
potential the effective dielectric constants $\varepsilon_{\perp}^{eff}$,
$\varepsilon_{\parallel}^{eff}$ are very \textit{different from the bulk
values} in $SrTiO_{3}$ (in $1ML$ $FeSe/TiO_{2}$ one has $\varepsilon\leq260$
\cite{Rebec-FeSe-TiO2}). This is physically plausible since for the nearest
(to the $FeSe$ monolayer) $TiO_{2}$ dipolar monolayer there is almost nothing
to screen in the direction perpendicular to $FeSe$, thus making $\varepsilon
_{\perp}^{eff}\ll$ $\varepsilon^{bulk}$. Note, that for the parameters assumed
in this analysis and for $T_{c0}\approx100$ $K$ one obtains rather large bare
pairing potential $V_{epi}^{0}\approx10$ $eV$ . This means that in the absence
of the $FSP$ in $EPI$ and for the density of states of the order $N(E_{F}%
)\sim0.5$ $(eV)^{-1}$ (typical for Fe-based superconductors) the bare coupling
constant $\lambda_{epi}^{0}=N(E_{F})V_{epi}^{0}$\ would be large,
$\lambda_{epi}^{0}\sim5$. We stress that the above theory is also applicable
to recently discovered $1ML$ $FeSe/TiO_{2}$ \cite{Rebec-FeSe-TiO2}. To
conclude, the high $T_{c0}$ in $1ML$ $FeSe/SrTiO_{3}$ (and $1ML$
$FeSe/TiO_{2}$) is obtained on the expense of the large maximal $EPI$ coupling
$V_{epi}^{0}$ which compensates smallness of the (detrimental) phase-volume
factor $(aq_{c})^{2}$.

\section{Effects of impurities on $T_{c0}$}

In clean systems with the $EPI-FSP$ mechanism of superconductivity $T_{c0}$ is
degenerate - it is equal in $s$- and $d$-channels. In the following we show,
that the $s$-wave superconductivity is also affected by isotropic non-magnetic
impurities, i.e. $T_{c0}$ is reduced and the Anderson theorem is violated.
This may have serious repercussions on the \textit{s-wave superconductivity}
in $1ML$ $FeSe/SrTiO_{3}$ (and $1ML$ $FeSe/TiO_{2}$) where $T_{c0}$ may depend
on chemistry. Then by using equation $Eq.$(\ref{Tc-s-wave}) from Appendix one
obtains $T_{c}^{(s)}$%
\[
\frac{T_{c}^{(s)}}{T_{c0}}=\frac{4}{\pi^{2}\rho}\psi(\frac{1}{2}+\frac{\rho
}{2})-\psi(\frac{1}{2}),
\]
where $\rho=\Gamma/\pi T_{c}$, $\Gamma=\pi n_{i}N(E_{F})u^{2}$, $n_{i}$ is the
impurity concentration and $u$ is the impurity potential. Let us consider some
limiting cases: ($1$) for $\Gamma\ll\pi T_{c}^{(s)}$one has $T_{c}%
^{(s)}\approx T_{c0}[1-7\zeta(3)\Gamma/\pi^{3}T_{c0}]$; ($2$) for $\Gamma
\gg\pi T_{c}$ one has $T_{c}^{(s)}\approx(\Gamma/2\pi)\exp(-\pi\Gamma
/4T_{c0})$, i.e. $T_{c}^{(s)}$ never vanishes. This means that in the
$EPI-FSP$ systems the non-magnetic impurity scattering is
\textit{pair-weakening} for the s-wave superconductivity.

In the case of \textit{d-wave superconductivity} the solution of
$Eq.$(\ref{Tc-d-wave}) in limiting cases is: \ ($1$) $T_{c}^{(d)}\approx
T_{c0}[1-2\Gamma/\pi T_{c0}]$ for $\Gamma\ll\pi T_{c}^{(d)}$. We point out
that the slope $-dT_{c}^{(d)}/d(\Gamma)=2/\pi$ is smaller than the slope for
the standard $d$-wave pairing, where $-dT_{c}/d(\Gamma)=\pi/4$. ($2$) \ For
$\Gamma>\Gamma_{cr}^{(FSP)}\approx(4/\pi)T_{c0}$ one has $T_{c}^{(d)}=0$, i.e.
the effect of non-magnetic impurities is \textit{pair-breaking}. Note, that
$\Gamma_{cr}^{(FSP)}>\Gamma_{cr}(=(2/\pi)T_{c0})$. These two results mean that
in the presence of non-magnetic impurities the $d$-wave superconductivity
which is due to the $EPI-FSP$ pairing is more robust than in the case of the
standard $d$-wave pairing. We stress, that the $T_{c}$ dependence on
non-magnetic impurities in $1ML$ $FeSe/SrTiO_{3}$ (as well as in$1ML$
$FeSe/TiO_{2}$) might be an important test for the $EPI-FSP$ pairing in this material.

Finally, it is worth of mentioning, that the real isotope effect in $T_{c0}$
of $1ML$ $FeSe/SrTiO_{3}$ (and in $1ML$ $FeSe/TiO_{2}$) might depend on the
type of non-magnetic impurities. If their potential is also long-ranged (for
instance due to oxygen deffects in the $TiO_{2}$ dipole layer), then there is
$FSP$ in the scattering potential, i.e. $u_{imp}^{2}(q)\approx u^{2}%
\delta(\mathbf{q})$. Then, such impurities affect in the same way $s$- and
$d$-wave pairing and they are pair weakening, as shown in
\cite{Dan-Dol-Kul-Oud}. Naimly, one has $(a)$ $T_{c}^{(s,d)}\approx
T_{c0}[1-4\Gamma_{F}/49T_{c0}]$ for $\Gamma_{F}\ll\pi T_{c}$, where
$\Gamma_{F}=\sqrt{n_{i}}u$; (b) $T_{c}^{(s,d)}\approx0.88$ $\Omega\exp
(-\pi\Gamma_{F}/4T_{c0})$, for $\Gamma_{F}\gg\pi T_{c}$. There are two
important results: $(1)$ There is a nonanalicity in $\Gamma_{F}\sim\sqrt
{n_{i}}$; $(2)$ there is a full isotope effect in the "dirty" limit
$\Gamma_{F}\gg\pi T_{c}$, i.e. $\alpha_{O}=1/2$, since $T_{c}^{(s,d)}%
\sim\Omega\sim M^{-1/2}$. We stress, that if the full isotope effect would be
realized experimentally in $1ML$ $FeSe/SrTiO_{3}$ (and in $1ML$ $FeSe/TiO_{2}%
$), then this does not automatically exclude the $FSP-EPI$ mechanism of
pairing, since it may be due to impurity effects. In that case the
nonanalicity of $\Gamma_{F}$ in $n_{i}$ might be a smoking-gun effect.

\section{Internal pair fluctuations reduce $T_{c0}$}

The $EPI-FSP$ theory, which predicts a long-range force between paired
electrons, opens a possibility for a pseudogap behavior in $1ML$
$FeSe/SrTiO_{3}$ (and $1ML$ $FeSe/TiO_{2}$). As we have discussed above, the
$EPI-FSP$ theory predicts a non-$BCS$ dependence of the critical temperature
$T_{c0}$, i.e. $T_{c0}=\left\langle V_{epi}(\mathbf{q})\right\rangle _{q}/4$.
However, this mean-field ($MFA$) value is inevitably reduced by the
\textit{phase} and \textit{internal Cooper pair fluctuations} - which are
present in systems with long-range attractive forces. Namely, in the $MFA$ the
order parameter $\Delta(\mathbf{x},\mathbf{\mathbf{x}}^{\prime}%
)(=V(\mathbf{x-\mathbf{x}}^{\prime})\langle\mathbf{\psi}_{\downarrow
}(\mathbf{x}^{\prime})\mathbf{\psi}_{\uparrow}(\mathbf{x})\rangle)$ depends on
the relative (internal) coordinate $\mathbf{r}=\mathbf{x}-\mathbf{\mathbf{x}%
}^{\prime}$ and the center of mass $\mathbf{R}=(\mathbf{x}+\mathbf{\mathbf{x}%
}^{\prime}\mathbf{\mathbf{)/2}}$, i.e. $\Delta(\mathbf{x},\mathbf{\mathbf{x}%
}^{\prime})=\Delta(\mathbf{r},\mathbf{\mathbf{R}})$. In usual superconductors
with short-range pairing potential one has $V_{sr}(\mathbf{x-\mathbf{x}%
}^{\prime})\approx V_{0}\delta(\mathbf{x-\mathbf{x}}^{\prime})$ and
$\Delta(\mathbf{r},\mathbf{\mathbf{R}})=\Delta(\mathbf{\mathbf{R}})$.
Therefore only the spatial ($\mathbf{\mathbf{R}}$-dependent) fluctuations of
the order parameter are important. In case of a long-range pairing potential
there are additional pair-fluctuations due to the dependence of $\Delta
(\mathbf{r},\mathbf{\mathbf{R}})$ on internal degrees of freedom (on
$\mathbf{r}$). The interesting problem of fluctuations in systems with
long-range attractive forces in $3D$ systems was studied in \cite{Yang} and we
sketch it briefly, because it shows that standard and $EPI-FSP$
superconductors belong to different universality classes.\ The best way to see
importance of the internal pair-fluctuations is to rewrite the pairing
Hamiltonian in terms of pseudospin operators (in this approximation first done
by P. Anderson the single particle excitations are not included)
\begin{equation}
\hat{H}=\sum_{\mathbf{k}\sigma}2\xi_{\mathbf{k}}\hat{S}_{\mathbf{k}\sigma}%
^{z}-(1/2)\sum_{\mathbf{k},\mathbf{k}^{\prime}}V_{\mathbf{k}-\mathbf{k}%
^{\prime}}(\hat{S}_{\mathbf{k}}^{+}\hat{S}_{\mathbf{k}^{\prime}}^{-}+\hat
{S}_{\mathbf{k}^{\prime}}^{+}\hat{S}_{\mathbf{k}}^{-}), \label{H-pseudo}%
\end{equation}
where $S_{\mathbf{k}\sigma}^{z}=(\hat{c}_{\mathbf{k}\uparrow}^{\dagger}\hat
{c}_{\mathbf{k}\uparrow}-\hat{c}_{-\mathbf{k}\downarrow}^{\dagger}\hat
{c}_{-\mathbf{k}\downarrow}-1)/2$, $\hat{S}_{\mathbf{k}\sigma}^{+}=\hat
{c}_{\mathbf{k}\uparrow}^{\dagger}\hat{c}_{-\mathbf{k}\downarrow}^{\dagger}%
$\ \cite{Yang}. This is a Heisenberg-like Hamiltonian in the momentum space.
In case of the $s$-wave superconductivity with short-range forces
$V_{sr}(\mathbf{x-\mathbf{x}}^{\prime})\approx V_{0}\delta
(\mathbf{x-\mathbf{x}}^{\prime})$ one has $V_{\mathbf{k}-\mathbf{k}^{\prime}%
}=const$ and the pairing potential is \textit{long-ranged in the momentum
space}. In that case it is justified to use \textit{the mean-field
approximation} $\hat{H}\rightarrow\hat{H}_{mf}=-\sum_{k}\mathbf{h}%
_{\mathbf{k}}\mathbf{\hat{S}}_{\mathbf{k}}$ with the mean-field $\mathbf{h}%
_{\mathbf{k}}=-2\xi_{\mathbf{k}}\mathbf{z+}\sum_{\mathbf{k}^{\prime}%
}V_{\mathbf{k}-\mathbf{k}^{\prime}}\langle S_{\mathbf{k}^{\prime}}%
^{x}\mathbf{x+}S_{\mathbf{k}^{\prime}}^{y}\mathbf{y\rangle}$ . The excitation
spectrum (with respect to the ground state) in this system have a gap, i.e.
$E(\mathbf{k})=2\sqrt{\xi_{\mathbf{k}}^{2}+\Delta_{\mathbf{k}}^{2}}$ where the
gap $\Delta_{\mathbf{k}}$ is the mean-field order parameter defined by
$\Delta_{\mathbf{k}}=\sum_{\mathbf{k}^{\prime}}V_{\mathbf{k}-\mathbf{k}%
^{\prime}}\langle\hat{S}_{\mathbf{k}^{\prime}}^{x}\rangle$. In case of the
$EPI-FSP$ pairing mechanism the pairing potential is long-ranged in real space
and short-ranged in the momentum space. For instance, in $1ML$ $FeSe/SrTiO_{3}%
$ (and $1ML$ $FeSe/TiO_{2}$) one has $V_{\mathbf{k}-\mathbf{k}^{\prime}}%
=V_{0}\exp\{-\left\vert \mathbf{k}-\mathbf{k}^{\prime}\right\vert /q_{c}\}$
with $q_{c}\ll k_{F}$, and the excitation spectrum is boson-like
$0<E(\mathbf{k})<2\sqrt{\xi_{\mathbf{k}}^{2}+\Delta_{\mathbf{k}}^{2}}$ (like
in the Heisenberg model) with large number of low-laying excitations (around
the ground state). This means, that there are many low-laying pairing states
above the ground-state in which pairs are sitting. This, so called internal
fluctuations effect, reduces $T_{c0}$ to $T_{c}$. For instance, tin $3D$
systems with $q_{c}\xi_{0}\ll1$ \cite{Yang} one has $T_{c}\approx(q_{c}\xi
_{0})T_{c0}$, where the coherence length $\xi_{0}=v_{F}/\pi\Delta_{0}$ and
$\Delta_{0}=2T_{c0}$. It is expected, that in the region $T_{c}<T<T_{c0}$ the
pseudogap ($PG$) phase is realized. However, in $2D$ systems, like $1ML$
$FeSe/SrTiO_{3}$ (and $1ML$ $FeSe/TiO_{2}$), there are additionally phase
fluctuations reducing $T_{c}$ further to the Berezinskii-Kosterliz-Thouless
value. We stress, that recent measurements of $T_{c}$ in $1ML$ $FeSe/SrTiO_{3}%
$ by the Meissner effect and resistivity ($\rho(T)$) give that $T_{c}^{(\rho
)}<T_{c}^{(M)}$ what may be partly due to these internal fluctuations of
Cooper pairs. It would be interesting to study theoretically these two kind of
fluctuations in $2D$ systems, such as $1ML$ $FeSe/SrTiO_{3}$ and $1ML$
$FeSe/TiO_{2}$.

\section{ Summary and Discussion}

In the paper we study the superconductivity in the $1ML$ $FeSe/SrTiO_{3}$ and
$1ML$ $FeSe/TiO_{2}$ sandwitch-structure, which contains one metallic $FeSe$
monolayer grown on the substrate $SrTiO_{3}$, or rutile $TiO_{2}^{(100)}$. It
turns out that in such a structure the \textit{Fermi surface is electron-like}
and the bands are pockets around the $M$-point in the Brillouin zone. The
bottom of the electron-like bands is around $(50-60)$ $meV$ below the Fermi
surface at $E_{F}$. The top of the hole-like band at the point $\Gamma$ lies
$80$ $meV$ below $E_{F}$ which means that pairing mechanisms based on the
electron-hole nesting are ruled out. This holds also for the pairing with
hole-incipient bands (very interesting proposal) \cite{Hirschfeld-review}. The
superconductivity in $1ML$ $FeSe/SrTiO_{3}$ and $1ML$ $FeSe/TiO_{2}$ is
realized in the $FeSe$ monolayer with $T_{c}\sim(60-100)$ $K$. The decisive
fact for making a theory is that the $ARPES$ spectra show s\textit{harp
replica bands} around $100$ $meV$ below the quasiparticle band, what is
approximately the energy of the oxygen optical phonon $\Omega\approx90$ $meV$.
The analysis of superconductivity is based on the semi-microscopic model -
first proposed in \cite{Lee-Interfacial-supplement}, \cite{Johnston1-2}, where
it is assumed that a $TiO_{2}$ dipolar layer is formed just near the
interface. In that model the oxygen vibrations create a dipolar electric
potential, which acts on electrons in the $FeSe$ monolayer, thus making the
$EPI$ interaction long-ranged. In the momentum space a forward scattering peak
($FSP$) appears, i.e. $EPI$ is peaked at small transfer momenta ($q<q_{c}\ll
k_{F}$) with $g_{epi}(\mathbf{q})=g_{0}\exp\{-q/q_{c}\}$. Here, this is called
the $EPI-FSP$\ pairing mechanism. The $EPI-FSP$ theory is formulated first in
\cite{Kulic-Zeyher1-2} for strongly correlated systems, while its extreme case
with delta-peak is elaborated in \cite{Dan-Dol-Kul-Oud} - see also
\cite{Kulic-PhysRep}. This limiting (delta-peak) case makes not only
analytical calculations easier, but it makes also a good fit to experimental
results \cite{Lee-Interfacial-supplement}, \cite{Johnston1-2}. In the
following, we summarize the main obtained results of the $EPI-FSP$ theory and
its relation to the $1ML$ $FeSe/SrTiO_{3}$ and $1ML$ $FeSe/TiO_{2}$ sandwitch-structures.

$(1)$ - The mean-field critical temperature $T_{c0}$ in the $s$-wave and
$d$-wave pairing channels is\textit{\ degenerate }and given by $T_{c0}%
=\left\langle V_{epi}(\mathbf{q})\right\rangle _{q}/4\approx(1/16\pi
)(aq_{c})^{2}V_{epi}^{0}$, where $V_{epi}(\mathbf{q})=2g_{epi}^{2}%
(\mathbf{q})/\hbar\Omega$ and the maximal pairing potential $V_{epi}%
^{0}(\equiv V_{epi}(q=0))=(2g_{0}^{2}/\Omega)$. On the first glance this
linear dependence of $T_{c0}$ on $V_{epi}^{(0)}$ seems to be favorable for
reaching high $T_{c0}$ - note in the $BCS$ theory $T_{c0}$ is exponentially
dependent on $V_{epi}^{(0)}$ and very small for small $N(E_{F})V_{epi}^{0}$.
However, for non-singular $g_{epi}(\mathbf{q})$ when $g_{epi}(\mathbf{q}=0)$
is finite, $T_{c0}$ is limited by the smallness of the phase-volume effect,
which is in $2D$ systems (such as $1ML$ $FeSe/SrTiO_{3}$ and $1ML$
$FeSe/TiO_{2}$) proportional to $(aq_{c})^{2}\ll1$. In that sense optimistic
claims that the $EPI-FSP$ mechanism leads inevitably to higher $T_{c0}$ - than
the one in the standard Migdal-Eliashberg theory, are not well founded. This
holds especially for $3D$ systems, where $T_{c0}\sim$ $(aq_{c})^{3}$ and
$T_{c0}^{(3d)}\ll T_{c0}^{(2d)}$ for the same value of $V_{epi}^{0}$. However,
higher $T_{c0}$ (with respect to to the $BCS$ case) can be reached by fine
tuning of $aq_{c}$ and $V_{epi}^{0}$. This is probably realized in $HTSC$
cuprates and with certainty in $1ML$ $FeSe/SrTiO_{3}$ and $1ML$ $FeSe/TiO_{2}%
$. The weak-coupling theory predicts the superconducting gap to be $\Delta
_{0}=2T_{c0}$ and for $T_{c0}\sim100$ $K$ one has $\Delta_{0}\sim16$ $meV$
what fits well the $ARPES$ experimental values
\cite{Lee-Interfacial-supplement}, \cite{Rebec-FeSe-TiO2}. Note, in order to
reach $T_{c0}=100$ $K$ for $aq_{c}\approx0.2$ a very large maximal
$EPI$\ coupling $V_{epi}^{0}\approx10$ $eV$ \ is necessary. For $N(E_{F}%
)\sim0.5$ $(eV)^{-1}$ the maximal coupling constant would be rather large,
i.e. $\lambda_{epi}^{0}(=N(E_{F})V_{epi}^{0})\approx5$. Note, that
$V_{epi}^{0}$ is almost as large as in the metallic hydrogen under high
pressure $p\sim20$ $Mbar$, where $T_{c0}\approx600$ $K$ with large $EPI$
coupling constant $\lambda_{epi}^{0}\approx7$ - this important prediction is
given in \cite{Maksimov-Savrasov}. In real $1ML$ $FeSe/SrTiO_{3}$ and $1ML$
$FeSe/TiO_{2}$ materials the contribution of another pairing mechanism,\ which
exists in the $FeSe$ film in absence of the substrate and is pronounced in the
\textit{s-wave channel} with $T_{c0}\approx8$ $K$, triggers the whole pairing
to be $s$-wave. The latter only moderately decreases the contribution of the
$EPI-FSP$ pairing mechanism. The existence of sharp replica bands in $1ML$
$FeSe/SrTiO_{3}$ and $1UCFeSe/TiO_{2}$ and large value of $V_{epi}^{0}$ imply
inevitably that the $EPI-FSP$ pairing mechanism is the main candidate to
explain superconductivity in these materials. We stress, that in $1ML$
$FeSe/SrTiO_{3}$ and $1ML$ $FeSe/TiO_{2}$ high $T_{c0}$ is obtained on the
expense of the large maximal $EPI$ coupling $V_{epi}^{0}$, which compensates
the small (detrimental) phase volume factor $(aq_{c})^{2}$.

$(2)$ - The semi-microscopic model proposed in
\cite{Lee-Interfacial-supplement}, \cite{Johnston1-2}, and refined slightly in
this paper, contains phenomenological parameters, such as $n_{d}$ - the number
of dipoles per unit cell, $q_{eff}$ - the effective dipole charge,
$\varepsilon_{\parallel}^{eff}$, $\varepsilon_{\perp}^{eff}$ - effective
parallel and perpendicular dielectric constanty in $SrTiO_{3}$ (and $TiO_{2}$)
near the interface, respectively. For $T_{c0}\approx100$ $K$ and by assuming
$aq_{c}\approx0.2,$ $q_{eff}\approx2e$, $n_{d}\sim2$ $/unit-cell$ makes
$\varepsilon_{\parallel}^{eff}\sim30$, $\varepsilon_{\perp}^{eff}\sim1$. These
values, which are physically plausible, are very far from $\varepsilon_{bulk}$
in the bulk $SrTiO_{3}$, where $\varepsilon_{bulk}\sim500-10^{4}$ (and
$\varepsilon_{bulk}\leq260$ in the rutile $TiO_{2}$). We point out that our
estimation of these parameters is based on the effective microscopic model
where the bulk $SrTiO_{3}$ is truncated by a monolayer ($1ML$) made of
$TiO_{2}$ \cite{Lee-Interfacial-supplement}. In reality it may happen that the
bulk $SrTiO_{3}$ is truncated by two monolayers ($2ML$) of $TiO_{2}$, as it is
claimed to be seen in the synchrotron $x$ray diffraction \cite{Zou-Bozovic}%
.This finding is confirmed by the $LDA$ calculations in \cite{Zou-Bozovic},
which show that for the $2ML$ $TiO_{2}$ structure: ($\emph{i}$) the electrons
are much easier transferred to the $FeSe$ metallic monolayer and ($\emph{ii}$)
the top of the hole band is shifted far below the electronic Fermi surface
than in the $1ML$ model. If the $2ML$ of $TiO_{2}$ is realized it could be
even more favorable for the $EPI-FSP$ pairing, since some parameters can be
changed in a favorable way. For instance, the effective charge could be
increased, i.e. $q_{eff}^{(2ML)}>q_{eff}^{(1ML)}$and since $T_{c0}\sim
q_{eff}^{2}$ the $2ML$ model may gives rise to higher critical temperature.

$(3)$ - The\textit{\ isotope effect }in $T_{c0}$\textit{\ }should be small
($\alpha_{O}\ll1/2$) since in leading order one has $T_{c0}\sim V_{epi}^{0}$,
where $V_{epi}^{0}$ is \textit{mass-independent}. This is contrary to
\cite{Johnston1-2} where $\alpha_{O}=1/2$. The next leading order gives
$\alpha_{O}\sim(T_{c0}/\Omega)<0.09$. We stress that the small isotope-effect
maybe a smoke-gun experiment for the $EPI-FSP$ pairing mechanism.

$(4)$ - In the $EPI$-$FSP$ pairing theory\textit{\ }the \textit{non-magnetic
impurities} affect both $s$-wave and $d$-wave pairing. In the case of
$s$-wave\textit{\ }they are\textit{\ pair-weakening}, while \textit{for }%
$d$-wave are\textit{\ pair-breaking}. However, the non-magnetic impurities
with forward scattering peak give in the "dirty" limit ($\Gamma_{F}\gg\pi
T_{c}$) the full isotope effect $\alpha_{O}=1/2$, since $T_{c}^{(s,d)}%
\sim\Omega\sim M^{-1/2}$. In that case, the nonanalicity of $T_{c}^{(s,d)}$
with respect to the impurity concentration $n_{i}$, would resolve the question
- what kind of pairing is realized in $1ML$ $FeSe/SrTiO_{3}$ and
$1UCFeSe/TiO_{2}$ - the $EPI-FSP$ or the standard $EPI$.

$(5)$ - In the case of the $EPI$-$FSP$ pairing the superconducting order
parameter depends strongly on the internal pair coordinate and of center of
mass, i.e. $\Delta=\Delta(\mathbf{r},\mathbf{R})$. The \textit{internal pair
fluctuations }reduce additionally the mean-field critical temperature so that
in the interval $T_{c}<T<T_{c0}$ a pseudogap behavior is expected.

$(6)$ - The \textit{EPI self-energy }in the normal state at $T=0$ and
$\xi(\mathbf{k}_{F})=0$ is given by $\Sigma_{epi}(\mathbf{k},\omega
)\approx-\lambda_{m}\omega/(1-(\omega/\Omega)^{2})$, where $\lambda
_{m}=\left\langle V_{epi}(\mathbf{q})\right\rangle _{q}/2\Omega$, which for
$G^{-1}(\mathbf{k},\omega)=0$ gives the dispersion energy of the quasiparticle
band $\omega_{1}=0$ and the replica bands $\omega_{2}$ and $\omega_{3}$. The
ratio of the ARPES intensities of the replica band $\omega_{2}$ and the
quasiparticle band $\omega_{1}$ at $T=0$ and at the Fermi surface ($k=k_{F}$)
is given by $R(T=0,k_{F})=(A_{2}/A_{1})=\lambda_{m}/2$. This means, that for
$\lambda_{m}\sim0.2$ the experimental value of $R(T=0,k_{F})$ should be
$(A_{2}/A_{1})\approx0.1$. This ratio is slightly smaller than the
experimental value $R(T\neq0,k=0)$ measured in
\cite{Lee-Interfacial-supplement}, \cite{Replica-bands}.

$(7$) Since the coupling constant $\lambda_{m}$ is \textit{mass-dependent},
$\lambda_{m}\sim M^{1/2}$ then the isotope effect in various quantities, in
$1ML$ $FeSe/SrTiO_{3}$ and $1ML$ $FeSe/TiO_{2}$ systems, may be a smoke-gun
experiment in favour of the $EPI-FPS$ theory. To remind the reader: $(i)$
$T_{c0}$ is almost \textit{mass-independent}; $(ii)$ the self-energy slope at
$\omega\ll\Omega$ is \textit{mass-dependent}, $(-d\Sigma/d\omega)\sim M^{1/2}%
$; $(iii)$ the $ARPES$ ratio $R$ of the replica band intensities is
\textit{mass-dependent}, $R\sim M^{1/2}$.

Concerning the role of $EPI$ in explaining superconductivity in $1ML$
$FeSe/SrTiO_{3}$ there were other interesting theoretical proposals. In
\cite{Gorkov1-2} the $EPI$ is due to the interaction with longitudinal optical
phonons and since $\Omega>E_{F}$ the problem is studied in anti-adiabatic
limit, where $T_{c}$ is also weakly dependent on the oxygen mass. In
\cite{Cohen} the substrate gives rise to an antiferromagnetic structure in
$FeSe$, which opens new channels in the $EPI$\ coupling in the $FeSe$
monolayer, thus giving rise for high $T_{c}$. In \cite{Zi} the intrinsic
pairing mechanism is assumed to be due to $J_{2}$-type spin fluctuations, or
antiferro orbital fluctuation, or nematic fluctuations. The extrinsic pairing
is assumed to be due to interface effects and the $EPI-FSP$ interaction. The
problem is studied by the sign-free Monte-Carlo simulations and it is found
that $EPI-FSP$ is an important ingredient for high $T_{c}$ superconductivity
in this system.

Finally, we would like to comment some possibilities for designing new and
complex structures based on $1ML$ $FeSe/SrTiO_{3}$ (or $1ML$ $FeSe/TiO_{2}$)
as a basic unit. The first nontrivial one is when a double-sandwich structure
with two interfaces is formed, i.e. $SrTiO_{3}/1ML$ $FeSe/SrTiO_{3}$ (or
$TiO_{2}/1ML$ $FeSe/TiO_{2}$). Naively thinking in the framework of the
$EPI-FSP$ pairing mechanism one expects in an "ideal" case doubling of
$T_{c0}$, since phonons at two interfaces are independent. However, this would
only happen when the electron-like bands on the Fermi surface due to the two
substrates were similar and if the condition $q_{c}v_{F}<\pi T_{c0}$ is kept
in order to deal with a sharp $FSP$. However, many complications in the
process of growing, such structures may drastically change properties, leading
even to a reduction of $T_{c0}$. It needs very delicate technology to control
the concentration of oxygen vacancies and appropriate charge transfer at both
interfaces. However, eventual solutions of these problems might give impetus
for superconductors with exotic properties. For instance, having in mind the
above exposed results on effects of non-magnetic impurities on $T_{c0}$, then
by controlling and manipulating their presence at both interfaces one can
design superconducting materials with wishful properties.

\textbf{Acknowledgment} M.L.K. is thankful to Rado\v{s} Gaji\'{c} for useful
discussions, comments and advises related to the experimental situation in the
field. M.L.K. highly appreciates fruitful discussions with Steve Johnston and
Yan Wang on ARPES\ of the replica bands at finite temperature, and on
microscopic parameters of the theory.

\section{Appendix}

\subsection{Migdal-Eliashberg equations in superconductors}

In the paper we study superconductivity with the $EPI-FSP$ mechanism of
pairing by including effects of non-magnetic impurities, too. The full set of
Migdal-Eliashberg equations is given for that case. The normal and anomalous
Green's functions are $G_{n}(\mathbf{k},\omega_{n})=$ $-[i\omega_{n}%
Z_{n}(\mathbf{k})+\bar{\xi}_{n}(\mathbf{k})]/D_{n}(\mathbf{k})$,
$F_{n}(\mathbf{k})=-Z_{n}(\mathbf{k})\Delta_{n}(\mathbf{k})/D_{n}(\mathbf{k}%
)$, respectively where $D_{n}(\mathbf{k})=[\omega_{n}Z_{n}(\mathbf{k}%
)]^{2}+\bar{\xi}_{n}^{2}(\mathbf{k})+[Z_{n}(\mathbf{k})\Delta_{n}%
(\mathbf{k})]^{2}$ ($\omega_{n}=\pi T(2n+1)$). Here, $Z_{n}(\mathbf{k})$ is
the wave-function renormalization defined by $i\omega_{n}(1-Z_{n}%
(\mathbf{k}))=(\Sigma(\mathbf{k},\omega_{n})-\Sigma(\mathbf{k},-\omega
_{n}))/2$, where the self-energy $\Sigma(\mathbf{k},\omega_{n})=\Sigma
_{epi}(\mathbf{k},\omega_{n})+\Sigma_{imp}(\mathbf{k},\omega_{n})$ describes
the $EPI$ and impurity scattering, respectively. The energy renormalization is
$\bar{\xi}_{n}(\mathbf{k})=\xi(\mathbf{k})+\chi_{n}(\mathbf{k})$, $\chi
_{n}(\mathbf{k})=(\Sigma(\mathbf{k},\omega_{n})+\Sigma(\mathbf{k},-\omega
_{n}))/2$ and $\Delta_{n}(\mathbf{k})$ is the superconducting order parameter.%
\begin{equation}
Z_{n}(\mathbf{k})=1+\frac{T}{\omega_{n}}\sum_{\mathbf{k}^{\prime},n^{\prime}%
}\frac{V_{eff}(n-n^{\prime},\mathbf{k}-\mathbf{k}^{\prime})\omega_{n^{\prime}%
}Z_{n^{\prime}}(\mathbf{k}^{\prime})}{D_{n^{\prime}}(\mathbf{k}^{\prime})}
\label{Z-nk}%
\end{equation}%
\begin{equation}
\bar{\xi}_{n}(\mathbf{k})=\xi(\mathbf{k})-T\sum_{\mathbf{k}^{\prime}%
,n^{\prime}}\frac{V_{eff}(n-n^{\prime},\mathbf{k}-\mathbf{k}^{\prime})\bar
{\xi}_{n^{\prime}}(\mathbf{k}^{\prime})}{D_{n^{\prime}}(\mathbf{k}^{\prime})}
\label{Ksi-nk}%
\end{equation}%
\begin{equation}
Z_{n}(\mathbf{k})\Delta_{n}(\mathbf{k})=T\sum_{\mathbf{k}^{\prime},n^{\prime}%
}\frac{V_{eff}(n-n^{\prime},\mathbf{k}-\mathbf{k}^{\prime})Z_{n^{\prime}%
}(\mathbf{k}^{\prime})\Delta_{n^{\prime}}(\mathbf{k}^{\prime})}{D_{n^{\prime}%
}(\mathbf{k}^{\prime})}, \label{Delta-nk}%
\end{equation}
where $V_{eff}(n-n^{\prime},\mathbf{k}-\mathbf{k}^{\prime})=V_{epi}%
(n-n^{\prime},\mathbf{k}-\mathbf{k}^{\prime})+V_{imp}(n-n^{\prime}%
,\mathbf{k}-\mathbf{k}^{\prime})$, $V_{epi}(n-n^{\prime},\mathbf{k}%
-\mathbf{k}^{\prime})=-g_{epi}^{2}(\mathbf{k}-\mathbf{k}^{\prime}%
)\mathcal{D}_{ph}(\mathbf{k}-\mathbf{k}^{\prime},\omega_{n}-\omega_{n^{\prime
}})$ and $V_{imp}(n-n^{\prime},\mathbf{k}-\mathbf{k}^{\prime})=\delta
_{nn^{\prime}}n_{imp}u^{2}(\mathbf{k}-\mathbf{k}^{\prime})/T$. Here, the
phonon Green's function in the Einstein model with the single frequency
$\Omega$ is given by $\mathcal{D}_{ph}(\mathbf{k}-\mathbf{k}^{\prime}%
,\omega_{n}-\omega_{n^{\prime}})=-2\Omega/(\Omega^{2}+(\omega_{n}%
-\omega_{n^{\prime}})^{2})$ while the impurity scattering is described in the
Born-approximation. Here, $n_{imp}$ is the impurity concentration and
$u(\mathbf{k}-\mathbf{k}^{\prime})$ is the impurity potential. To these three
equations one should add the equation for the chemical potential $\mu$, i.e.
$N=\sum G_{n}(\mathbf{k},\omega_{n};\mu)=const$. However, in the following we
study only problems where the (small) change of $\mu$ due to $EPI$ and
impurity scattering does not change the physics of the problem. For instance
we do not study problems such as $BCS-BEC$ transition, where the equation for
$\mu$ plays important role, etc.

Note, that in the case of systems with very \textit{large Fermi energy}
$E_{F}$ and with an \textit{isotropic} $EPI$ ($Z_{n}(\mathbf{k})\equiv Z_{n}$,
$\bar{\xi}_{n}(\mathbf{k})\rightarrow0$) one integrates over the energy
$\xi_{\mathbf{k}^{\prime}}$ by introducing the density of states at the Fermi
surface $N(0)$, i.e. $\sum_{\mathbf{k}^{\prime}}(...)\Rightarrow
N(0)\int_{-\infty}^{\infty}(...)d\xi_{\mathbf{k}^{\prime}}$. This leads to
standard Migdal-Eliashberg equations.%
\begin{equation}
Z_{n}=1+\frac{\pi T}{\omega_{n}}\sum_{n^{\prime}}\frac{N(0)V_{eff}%
(n-n^{\prime})\omega_{n^{\prime}}Z_{n^{\prime}}}{\sqrt{(\omega_{n^{\prime}%
}Z_{n^{\prime}})^{2}+\Delta_{n^{\prime}}^{2}}} \label{Zn}%
\end{equation}%
\begin{equation}
Z_{n}\Delta_{n}=\pi T\sum_{n^{\prime}}\frac{N(0)V_{eff}(n-n^{\prime
},Z_{n^{\prime}}\Delta_{n^{\prime}}}{\sqrt{(\omega_{n^{\prime}}Z_{n^{\prime}%
})^{2}+\Delta_{n^{\prime}}^{2}}} \label{Delta-n}%
\end{equation}
In the case of \textit{strongly momentum-dependent} $EPI-FSP$,$\ $where
$V_{epi}(n-n^{\prime},\mathbf{q})$ is finite for $\left\vert \mathbf{q}%
\right\vert <q_{c}\ll k_{F}$, the Migdal-Eliashberg equations are given by%
\begin{equation}
Z_{n}(\xi)=1+\frac{T}{\omega_{n}}\sum_{m}\left\langle V_{epi}(n-m,\mathbf{q}%
)\right\rangle _{q}\frac{\omega_{m}Z_{m}(\xi)}{D_{m}(\xi)}, \label{Zn-FSP}%
\end{equation}

\begin{equation}
\bar{\xi}_{n}(\xi)=\xi(\vec{k})-T\sum_{m}\frac{\left\langle V_{epi}%
(n-m,\mathbf{q})\right\rangle _{q}}{D_{m}(\xi)}\bar{\xi}_{m}(\xi),
\label{Ksi-FSP}%
\end{equation}%
\begin{equation}
Z_{n}(\xi)\Delta_{n}(\xi)=T\sum_{m}\left\langle V_{epi}(n-m,\mathbf{q}%
)\right\rangle _{q}\frac{Z_{m}(\xi)\Delta_{m}(\xi)}{D_{m}(\xi)}.
\label{Delta-n-FSP}%
\end{equation}

\subsection{Effects of non-magnetic impurities on $T_{c0}$ in the $EPI-FSP$
theory}

In this paper we study the superconductivity which is due to $EPI-FSP$ of the
Einstein phonon with $\Omega^{2}\ll(2\pi T_{c0})^{2}$. In that case
$\left\langle V_{epi}(n-m,\mathbf{q})\right\rangle _{q}\approx\left\langle
V_{epi}(0,\mathbf{q})\right\rangle _{q}=\left\langle 2g_{epi}^{2}%
(\mathbf{q})\right\rangle _{q}/\Omega$ and the contribution to $Z_{n}(\xi)$ is
$\sim\lambda_{m}=\left\langle V_{epi}(0,\mathbf{q})\right\rangle _{q}/2\Omega
$. Since in the weak coupling limit one has $\lambda_{m}\ll1$ then we neglect
this contribution. Also the non-Migdal corrections can be neglected in this
case - see \cite{Johnston1-2} The effects of non-magnetic impurities on
$T_{c0}$ is studied in the standard model with weakly momentum dependent
impurity potential $u(\mathbf{k}-\mathbf{k}^{\prime})\approx const$ . In that
case $Z_{n}(\xi)$ contains the impurity term only. After the integration of
the impurity part over the energy $\xi^{\prime}$ in $Eqs.$(\ref{Z-nk}%
-\ref{Delta-nk}) -see \cite{Allen-Mitrovic}, and putting $\xi=0$ (since in
that case $\Delta_{n}(\xi=0)$ is maximal) one obtains ($D_{m}(\xi
)\approx\omega_{n}^{2}Z_{n}^{2}$) for the \textit{s-wave pairing}
($\Delta=const$)
\begin{equation}
Z_{n}=1+\frac{\Gamma}{\left\vert \omega_{n}\right\vert } \label{Zn-imp}%
\end{equation}%
\begin{equation}
Z_{n}\Delta_{n}=T_{c}\sum_{m}\left\langle V_{epi}(n-m,\mathbf{q})\right\rangle
_{q}\frac{Z_{m}\Delta_{m}}{D_{m}(0)}+\frac{\Gamma}{\left\vert \omega
_{n}\right\vert }\Delta_{n}. \label{Delta-n-imp}%
\end{equation}
Note, the the second term on the right side cancels the same term on the left
side. In the approximation $\Delta_{n}(\xi)\approx\Delta$ one obtains the
equation for impurity dependence of $T_{c}^{(s)}(\Gamma)$ for a \textit{s-wave
superconductor}
\begin{equation}
1=T_{c}^{(s)}\left\langle V_{epi}(0,\mathbf{q})\right\rangle _{q}\sum_{m}%
\frac{1}{\omega_{n}^{2}Z_{n}}. \label{Tc-s-wave}%
\end{equation}
We point out that in the case of \textit{d-wave superconductivity}
$\Delta=\Delta(\varphi)$ is angle dependent on the Fermi surface and changes
sign. In that case the last term in $Eq.$(\ref{Delta-n-imp}) $\Delta$ should
be replaced by $\left\langle \Delta(\varphi\right\rangle )=0$ giving equation
for $T_{c}^{(d)}$%
\begin{equation}
1=T_{c}^{(d)}\left\langle V_{epi}(0,\mathbf{q})\right\rangle _{q}\sum_{m}%
\frac{1}{\omega_{n}^{2}Z_{n}^{2}}. \label{Tc-d-wave}%
\end{equation}
Note, $Z_{n}$ vs $Z_{n}^{2}$ renormalization for the $s$-wave and $d$-wave
superconductivity, respectively.

\subsection{$EPI-FSP$ self-energy in the normal state}

We shall calculate the self-energy at $T=0$. The leading order self-energy (on
the Matsubara axis) in the Migdal-Eliashberg theory of $EPI$ is given by
\begin{equation}
\Sigma_{epi}(\mathbf{k},\omega_{n})=-T\sum_{\mathbf{q},\Omega_{m}}g_{epi}%
^{2}(\mathbf{q})\mathcal{D}_{ph}(\mathbf{q},\Omega_{m})G(\mathbf{k}%
+\mathbf{q},\omega_{n}-\Omega_{m}), \label{Sigma-ME}%
\end{equation}
where $\omega_{n}=\pi T(2n+1)$ and $\Omega_{m}=2\pi mT$, $\mathcal{D}%
_{ph}(\mathbf{q},\Omega_{m})=-2\Omega/(\Omega^{2}+\Omega_{m}^{2})$,
$G(\mathbf{k},\omega_{n})=1/(i\omega_{n}-\xi_{\mathbf{k}})$. By defining
$V_{epi}(\mathbf{q},0)=2g_{epi}^{2}(\mathbf{q})/\Omega$ \ and after summation
over $\Omega_{m}$ in $Eq.$(\ref{Sigma-ME}) one obtains (note that
$V_{epi}(\mathbf{q},0)=V_{epi}(-\mathbf{q},0)$
\begin{align}
\Sigma_{epi}(\mathbf{k},\omega_{n})  &  =\frac{\Omega}{2}\sum_{\mathbf{q}%
}V_{epi}(\mathbf{q},0)\label{Sigma-integ}\\
&  \times\left[  \frac{n_{F}(\xi_{\mathbf{k+q}})}{i\omega_{n}-\xi
_{\mathbf{k+q}}+\Omega}+\frac{1-n_{F}(\xi_{\mathbf{k+q}})}{i\omega_{n}%
-\xi_{\mathbf{k+q}}-\Omega}\right]  .\nonumber
\end{align}
Let us calculate $\Sigma_{epi}$ at $\mathbf{k}_{F}$. Since $\mathbf{qv}%
_{F}=qv_{F}\cos\theta$ and by taking into account that $n_{F}(\xi
_{\mathbf{k}_{F}\mathbf{+q}})=1$ for $\cos\theta<0$, $n_{F}(\xi_{\mathbf{k}%
_{F}\mathbf{+q}})=0$ for $\cos\theta>0$ one obtains for $q_{c}v_{F}\ll\Omega$
\begin{equation}
\Sigma_{epi}(\mathbf{k},\omega_{n})=-\lambda_{m}\frac{i\omega_{n}}%
{1-(\frac{i\omega_{n}}{\Omega})^{2}}, \label{Sigma-final}%
\end{equation}
where $\lambda_{m}=\left\langle V_{epi}(\mathbf{q})\right\rangle _{q}/2\Omega$
and $\left\langle V_{epi}(\mathbf{q})\right\rangle _{q}=Ns_{c}(2\pi)^{-2}\int
d^{2}qV_{epi}(\mathbf{q},0)$, $s_{c}$ is the surface of the $FeSe$ unit cell.
Note, that $V_{epi}(\mathbf{q},0)=2g_{epi}^{2}(\mathbf{q})/\Omega$ and
$g_{epi}(\mathbf{q})(=(g_{0}/\sqrt{N})e^{-q/q_{c}})$ so that $N$ disappears
from $\lambda_{m}(=\left\langle V_{epi}(\mathbf{q})\right\rangle _{q}%
/2\Omega)$.

\end{document}